# Measurement of breast-tissue x-ray attenuation by spectral mammography: first results on cyst fluid

Erik Fredenberg,[1] David R Dance,[2,3] Paula Willsher,[4] Elin Moa,[1] Miriam von Tiedemann,[1] Kenneth C Young[2,3] and Matthew G Wallis[4]

[1] Philips Mammography Solutions, Smidesvägen 5, 171 41 Solna, Sweden
[2] NCCPM, Royal Surrey County Hospital, Guildford GU2 7XX, United Kingdom
[3] Department of Physics, University of Surrey, Guildford GU2 7XH, United Kingdom
[4] Cambridge Breast Unit and NIHR Cambridge Biomedical Research Centre, Addenbrookes Hospital, Hills Road, Cambridge CB2 0QQ, United Kingdom

E-mail: erik.fredenberg@philips.com

**Abstract**
Knowledge of x-ray attenuation is essential for developing and evaluating x-ray imaging technologies. For instance, techniques to better characterize cysts at mammography screening would be highly desirable to reduce recalls, but the development is hampered by the lack of attenuation data for cysts. We have developed a method to measure x-ray attenuation of tissue samples using a prototype photon-counting spectral mammography unit. The method was applied to measure the attenuation of 50 samples of breast cyst fluid and 50 samples of water. Spectral (energy-resolved) images of the samples were acquired and the image signal was mapped to equivalent thicknesses of two known reference materials, which can be used to derive the x-ray attenuation as a function of energy. The attenuation of cyst fluid was found to be significantly different from water. There was a relatively large natural spread between different samples of cyst fluid, whereas the homogeneity of each individual sample was found to be good; the variation within samples did not reach above the quantum noise floor. The spectral method proved stable between several measurements on the same sample. Further, chemical analysis and elemental attenuation calculation were used to validate the spectral measurement on a subset of the samples. The two methods agreed within the precision of the elemental attenuation calculation over the mammographic energy range.

## 1. Introduction

Basic knowledge of the x-ray attenuation of tissue is essential for the development of new x-ray imaging technologies, as well as the study of the performance of existing technologies. In the field of breast imaging, only a few measurements have been reported of the attenuation of healthy breast tissue (Hammerstein *et al*, 1979, Johns and Yaffe, 1987, Carroll *et al*, 1994, Chen *et al*, 2010, Tomal *et al*, 2010), of breast tumors (Johns and Yaffe, 1987, Carroll *et al*, 1994, Chen *et al*, 2010, Tomal *et al*, 2010), and of fibroadenomas (Tomal *et al*, 2010). To our knowledge, attenuation data for breast cyst fluid have not been reported previously, although cysts are among the most common mammographic findings.

  Knowledge of the x-ray attenuation of tissue is particularly important for the development of new applications of unenhanced spectral imaging. Unenhanced spectral imaging is an emerging x-ray imaging technology that measures tissue properties, without injection of a contrast agent, using differences in attenuation as a function of energy. Unenhanced spectral imaging has been employed in mammography to improve the image signal-to-noise ratio (Tapiovaara and Wagner, 1985, Cahn *et al*, 1999), improve lesion visibility (Johns and Yaffe, 1985, Kappadath and Shaw, 2008, Taibi *et al*, 2003, Fredenberg *et al*, 2010b), and to measure breast density (Ding and Molloi, 2012). Another potential application is to distinguish cysts from tumors at screening (Norell *et al*, 2012). Round lesions are one of the commonest mammographic findings and it is difficult to determine whether they are cystic or solid, particularly when the margin is partly obscured. Techniques to improve characterization at screening would therefore be highly desirable to reduce recalls, but the development of x-ray techniques for this purpose is hampered because of a lack of attenuation data.

We have developed a method to measure the energy-dependent x-ray attenuation of tissue samples using spectral imaging. The method is based on mapping of the attenuation to equivalent thicknesses of two reference materials. An advanced prototype clinical spectral-imaging mammography system was used together with the method to measure the attenuation of cyst fluid. Preliminary results of this study have been presented by Fredenberg *et al* (2013).

## 2. Materials and Methods

*2.1. Spectral mammography system*

The Philips MicroDose Mammography system comprises a tungsten-target x-ray tube with aluminum filtration, a pre-collimator, and an image receptor, which is scanned across the object (Figure 1, left). The image receptor consists of photon-counting silicon strip detectors with corresponding slits in the pre-collimator (Figure 1, right). This multi-slit geometry rejects virtually all scattered radiation (Åslund *et al*, 2006). 32 kV acceleration voltage was used for all measurements presented in this study.

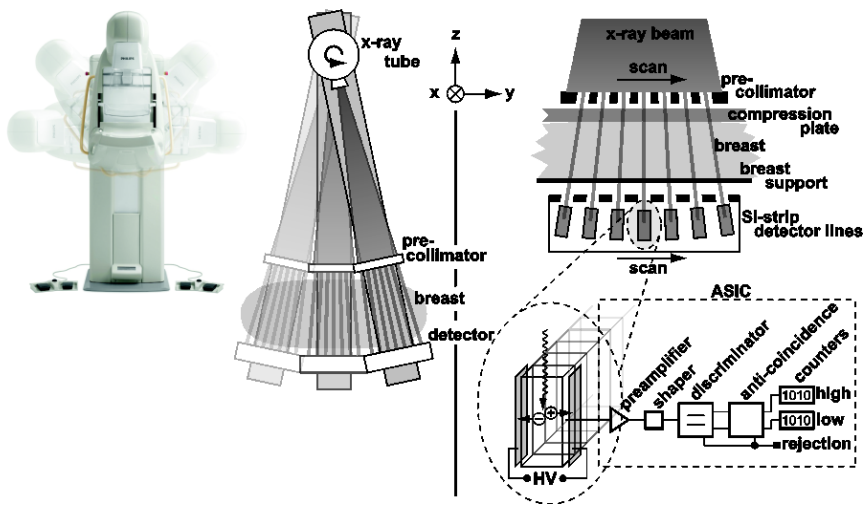

**Figure 1: Left:** Photograph and schematic of the Philips MicroDose Mammography system. **Right:** The spectral image receptor and electronics.

Photons that interact in the detector are converted to pulses with amplitude proportional to the photon energy. A low-energy threshold rejects virtually all pulses below a few keV, which are considered to be generated by noise. As we have reported previously, prototype MicroDose systems for spectral imaging have been developed (Fredenberg *et al*, 2010a). The main difference compared to the conventional system is the addition of a high-energy threshold to the detector, which sorts the detected pulses into two bins according to energy. The threshold was set close to 20 keV, which yields approximately equal number of counts in each bin for a typical transmitted spectrum. The energy resolution at this energy level is approximately 5 keV full width at half maximum (Fredenberg *et al*, 2010a).

*2.2. Background of spectral imaging*

For most natural body constituents at mammographic x-ray energies, it is fair to ignore absorption edges. X-ray attenuation is then made up of only two interaction effects, namely photoelectric absorption and scattering processes (Alvarez and Macovski, 1976, Lehmann *et al*, 1981, Johns and Yaffe, 1985). Accordingly, in the mammographic energy range, a linear combination of any two materials of different and low atomic number can approximately simulate the energy-dependent attenuation of a third material of a given thickness,

$$t_{\text{sample}} \times \mu_{\text{sample}}(E) = t_1 \times \mu_1(E) + t_2 \times \mu_2(E). \tag{1}$$

We call these materials reference materials, and if this relationship is assumed to hold exactly, then the associated normalized reference thicknesses $[t_1, t_2]/t_{\text{sample}}$ are unique descriptors of the energy dependent sample attenuation ($\mu_{\text{sample}}$) given the known attenuations of the reference materials ($\mu_1$ and $\mu_2$). Further, and with the same assumption, the detected signal ($I$) in a photon-counting x-ray detector would be identical for a tissue sample and for the equivalent combination of reference materials, regardless of incident energy spectrum ($\Phi(E)$) or detector response ($\Gamma(E)$),

$$\begin{aligned} I_{\text{sample}} &= I_0 \times \int \exp\bigl(-\mu_{\text{sample}}(E) \times t_{\text{sample}}\bigr) \times \Phi(E) \times \Gamma(E) \times \mathrm{d}E \\ &= I_0 \times \int \exp(-\mu_1(E) \times t_1 - \mu_2(E) \times t_2) \times \Phi(E) \times \Gamma(E) \times \mathrm{d}E = I_{\text{reference}}. \end{aligned} \tag{2}$$

Hence, measurements of attenuation at two different energies (or for two different energy spectra) yield a non-linear system of equations, which, for known $t_{\text{sample}}$, can be solved for $t_1$ and $t_2$. Measurements at more than two energies yield an over-determined system of equations under the assumption of only two independent interaction processes, and would, in principle, be redundant. Equations (1) and (2) assume that scattering processes can be treated as absorption, which is true only for x-ray detector geometries with efficient scatter rejection, such as multi-slit (Åslund *et al*, 2006).

In this work aluminum (Al) and polymethyl methacrylate (PMMA) have been used as the two reference materials. In common with previous studies (Alvarez and Macovski, 1976), it is useful to express the equivalent Al and PMMA thicknesses in terms of the corresponding Al-PMMA vector with magnitude and angle given by

$$r = \sqrt{t_1^2 + t_2^2} \qquad \text{and} \qquad \theta = \tan\left(\frac{t_1}{t_2}\right). \tag{3}$$

The magnitude $r$ is directly proportional to the thickness of the sample, whereas the angle $\theta$ is related to the attenuation energy dependence and is independent of sample thickness.

*2.3. Spectral attenuation measurements*

We have developed a method to measure x-ray attenuation of tissue samples by solving Eq. (2) for $t_1$ and $t_2$ in spectral images acquired with the Philips MicroDose system. Liquid samples were placed in PMMA cuvettes, and imaged with the mammography stand in vertical mode (Figure 2). An Al and PMMA step wedge was positioned adjacent to the sample and was present in each image to provide a reference grid of thickness/material combinations for comparison to the sample. X-ray attenuation was measured by mapping the high- and low energy counts obtained from a region-of-interest (ROI) located on the sample ($I_{\text{sample}}$) against those obtained from ROIs on the step wedge ($I_{\text{reference}}(t_{\text{Al}}, t_{\text{PMMA}})$). Linear Delaunay triangulation in the log domain was used to find intermediate reference values. The cuvettes were specified to be approximately 1-cm deep, but the precise thickness ($t_{\text{sample}}$) was measured with high accuracy in order to produce normalized values of $t_{\text{Al}}$ and $t_{\text{PMMA}}$.

The x-ray attenuation of 50 samples of cyst fluid was measured. Aspiration of the cyst fluid was part of routine clinical examination at the Cambridge Breast Unit, and the samples were used in the study after obtaining informed consent. A sample of distilled water was imaged adjacent to the cyst fluid in each image for reference.

Four images were acquired of each sample in order to estimate random fluctuations between measurements, referred to below as inter-image variability. In each image, data from four ROI locations on the sample were acquired in order to estimate inhomogeneities within the sample and image, referred to below as intra-image variability. Data from all ROI locations and images were added in order to estimate the expectation value for each sample. The spread between the expectation values for different samples is denoted total variability. It is likely that the major part of the total variability is inter-sample variability, but this will also include inter- and intra-image variability.

All of the sample and image variability measures are overlaid with quantum noise. The fundamental quantum noise level was estimated by assuming Poisson statistics for the number of detected photons and error propagation through the interpolation process. Two-sample *t*-tests and chi-square variance tests were used to test the differences between the means and the variances of the different measurements.

The exact thicknesses of the PMMA step wedge components were measured with a micrometer screw, and the PMMA density was determined by weighing the step wedge with high-precision scales. The cuvette wall thickness, container depth, and the variation between cuvettes were determined by measuring the outer dimensions and wall thicknesses on 4 cuvettes using a micrometer screw. All cuvettes were from the same molding batch so variations can be expected to be minimal. The thickness of the PMMA mounting plates were measured at the step wedge and sample locations to take into account any variation in thickness. The PMMA densities of the mounting plates were not measured because the difference in thickness between sample and step wedge locations is small so any reasonable deviation from a standard value would have minimal impact. The PMMA density of the cuvette was also not measured as the cuvette wall was thin compared to the sample thickness.

Errors in the PMMA thicknesses were estimated as the standard deviation of measurements at several locations, or as the precision of the micrometer screw, whichever was largest. Weighing errors were estimated by repeated measurements. Density errors were found by propagating weighing and thickness errors, or, in case the density was not measured, as the approximate variation of densities in off-the-shelf PMMA from different manufacturers. The effect of uncertainties in sample thickness is a scaling of $r$ (cf. Eq. (3)).

The Al steps in the step wedge were constructed from layers of Al foil (purity 99.999%). The thicknesses of these were determined by measuring the area and weight of an approximately 100×100 mm$^2$ Al sheet, assuming the elemental density, and solving for the thickness. The error in Al thickness was estimated as the standard deviation of measurements with a micrometer screw at several locations on the Al sheet.

Put together, the systematic error caused by uncertainties in thickness and density measurements was estimated to be within 43 μm PMMA and 1 μm Al (corresponding to errors of less than 0.5% for a 10 mm water sample). The random error caused by variation between cuvettes was found to be approximately one order of magnitude lower: 3 μm PMMA and 0.1 μm Al.

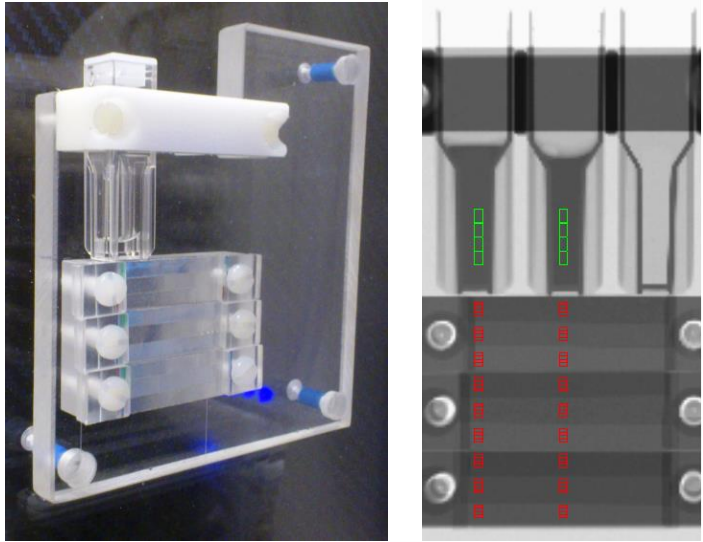

**Figure 2: Left:** Photograph of the Al-PMMA step wedge and the sample holder with one cuvette in place. **Right:** X-ray image of the sample holder and step wedge with three cuvettes in place. The center-most cuvette is filled with cyst fluid, and the left one is filled with distilled water for reference. The right-most cuvette is empty and was not used for measurements in this study. Sample ROIs that were used for the evaluation are indicated in green and step-wedge ROIs in red. There were four independent measurement points on each sample, which are illustrated by gridded ROIs.

*2.4. Validation of the spectral measurement method*

To validate the spectral attenuation measurement procedure, the linear attenuation coefficient and the equivalent thicknesses of reference materials were calculated from the elemental compositions of cyst fluid and water. The composition of cyst fluid was determined by sending five of the samples for quantitative elemental chemical analysis by Warwick Analytical Services (Warwick, UK). The percentage compositions by weight of elements H, C and N were determined with an accuracy of 0.3% (absolute) and elements S, Ca, K, Na, P and Cl with a relative accuracy of 1%. The amount of O present was determined by subtraction of the compositions of the aforementioned elements from 100%. Two analyses were made for each sample. The distilled water samples were assumed pure with corresponding elemental composition.

The density of each cyst sample and of water was measured by weighing $5.0\ \text{cm}^3$ of the fluid (estimated random measurement uncertainty 0.3%). In order to correct for a small systematic error, the results were scaled so that the density of water matched the tabulated value at 21°C (0.57% scaling on average). Further, the spectral measurements were performed at a slightly different temperature (23°C room temperature) and the cyst fluid density was scaled with the same factor as the tabulated one for water to account for the lower density at the higher temperature (scaling of <0.1%).

The linear attenuation coefficient can be calculated from the measured elemental composition and density according to

$$\mu_{\text{sample}}(E) = \rho_{\text{sample}} \times \sum \mu_i(E)/\rho_i \times w_i, \qquad (4)$$

where $\mu_i/\rho_i$ is the published mass attenuation coefficient of element $i$ (Berger *et al*, 2005), and $w_i$ is the corresponding fraction by weight. In principle, the linear attenuation coefficient could then be related to reference thicknesses $t_1$ and $t_2$ for Al and PMMA by calculating Eq. (1) and Eq. (4) at two arbitrary energies and solving the resulting linear system of equations. However, the limited validity of the assumptions underpinning Eq. (1) may lead to a slight dependence on the energies chosen. Instead, we used a linear least-squares fit over the x-ray spectrum energy bins ($E_1 \ldots E_N$), weighted with the expected spectrum after the sample ($\phi(E)$) and the detector response ($\Gamma(E)$). This procedure corresponds to the following minimization with respect to $t_1$ and $t_2$:

$$\min_{t_1,t_2} \sum_{n=1}^{N} \phi(E_n) \times \Gamma(E_n) \times \left(\mu_1(E_n) \times t_1 + \mu_2(E_n) \times t_2 - \mu_{\text{sample}}(E_n) \times t_{\text{sample}}\right)^2. \qquad (5)$$

Weighting with the spectrum is a second-order correction, which was found to influence the calculated reference thicknesses in the order of one percent compared to equal weighting at all energies. A relatively simple model without experimental validation was therefore deemed accurate enough to calculate $\phi$ and $\Gamma$. For $\phi$, we used a generic tungsten spectrum model (Boone *et al*, 1997), simply attenuated by 0.5 mm Al and the appropriate amount of PMMA from the sample holder. In $\Gamma$, all detector effects were ignored except for quantum efficiency, which was calculated by assuming a 3-mm-thick silicon detector.

Further, the published linear attenuation coefficients are associated with errors compared to experimental data typically in the few percent range (Saloman *et al*, 1988). There are indications that deviations increase towards lower energies (NIST, 1998), and therefore only energies $\geq 20$ keV were included in the validation.

## 3. Results and discussion

### 3.1. Validation of the spectral measurement method

The elemental composition of cyst fluid determined by chemical analysis is shown in Table 1 as the mean of all samples. The means of the two measurements of each sample differed by no more than 0.15% from the individual readings for H, C, N and O and by no more than 0.04% for the remaining elements. Also shown in Table 1 is the total variability between samples.

**Table 1:** Results from the analysis of the elemental composition of cyst fluid. The composition is given in percent weight. The total variability is given as one standard deviation. The composition of water is shown for reference. The density is for a temperature of 23°C.

|   |   | cyst mean | cyst total variability | water |
|---|---|---|---|---|
| O | [%] | 83.8 | 1.47 | 87.1 |
| H | [%] | 11.0 | 0.144 | 12.9 |
| C | [%] | 3.84 | 1.09 | - |
| N | [%] | 0.635 | 0.402 | - |
| K | [%] | 0.28 | 0.17 | - |
| Na | [%] | 0.20 | 0.11 | - |
| Cl | [%] | 0.14 | 0.13 | - |
| S | [%] | 0.080 | 0.039 | - |
| P | [%] | 0.011 | 0.004 | - |
| Ca | [%] | 0.009 | 0.003 | - |
| density | [g/cm$^{-1}$] | 1.013 | 0.007 | 0.998 |

The equivalent Al and PMMA thicknesses calculated from the elemental composition and measured by the spectral method on the same samples are listed in Table 2 together with the total variability. There were relatively large differences between the mean thickness values determined by the two different methods, and for water the difference was significant at the 1% significance level (determined by a one-sample *t*-test). The differences in mean between cyst fluid and water ($\Delta$), also listed in Table 2, were, however, smaller and not in any case significantly different for the two methods. As $\Delta$ is mainly sensitive to relative and random errors, the good agreement in $\Delta$ indicates that the discrepancies in mean values are to a large extent systematic. Table 2 also shows the angle ($\theta$) and magnitude ($r$) of the Al-PMMA vector. These values are mainly for illustration as they are not independent of the Al and PMMA thicknesses and therefore exhibit the same behavior: There are significant discrepancies in the mean values, whereas the $\Delta$ values were not significantly different.

The systematic errors in the mean values may be partly caused by errors in the density and thickness measurements, but the deviations are substantially larger than expected from the error analysis in Sec. 2.3. A major part of the systematic discrepancy is therefore expected to come from errors in the linear attenuation values used to calculate the PMMA and Al thicknesses from the elemental composition. Firstly, the coherent and incoherent scattering cross sections for compounds are not exactly the weighted sums of the cross sections for individual elements and Eq. (4) can only be considered an approximation. Further, Saloman *et al* (1988) report that errors in the elemental cross section tabulations are typically in the percentage range and hence not negligible.

Figure 3 shows the ratio of the linear attenuation coefficients for cyst fluid as determined by the spectral measurement and by the elemental composition. The mean absolute difference between the spectral and elemental composition methods for cyst fluid over the 20-40 keV range was 0.27% and the maximum was 0.35%, which is well within the expected errors of the cross section tabulations. We therefore conclude that the spectral measurement method has at least as good accuracy as calculation from the elemental composition.

**Table 2:** Validation of the spectral measurement. Equivalent PMMA and Al thicknesses for a 10-mm sample, calculated from the elemental composition of water and 5 samples of cyst fluid, and measured with the spectral method on the same samples. The total variability is quantified with the standard deviation. The separation of cyst fluid and water is presented as the difference between the means ($\Delta$). Additionally, the angle ($\theta$) and magnitude ($r$) of the Al-PMMA vectors are given and marked in grey to indicate that these measures are not independent of the PMMA and Al thicknesses.

|  |  | spectral |  | elemental |  |
|---|---|---|---|---|---|
|  |  | mean | total variability | mean | total variability |
| Cyst: | PMMA [mm]: | 8.12 | 0.78% | 8.02 | 0.76% |
|  | Al [mm]: | 0.320 | 3.2% | 0.328 | 3.4% |
|  | $\theta$ [mrad]: | 39.5 | 3.3% | 40.9 | 3.5% |
|  | $r$ [mm]: | 8.13 | 0.78% | 8.03 | 0.76% |
| Water: | PMMA [mm]: | 8.01 | 0.30% | 7.96 | - |
|  | Al [mm]: | 0.274 | 0.80% | 0.286 | - |
|  | $\theta$ [mrad]: | 34.3 | 0.87% | 36.0 | - |
|  | $r$ [mm]: | 8.01 | 0.30% | 7.96 | - |
| $\Delta$: | PMMA [mm]: | 0.12 | - | 0.06 | - |
|  | Al [mm]: | 0.046 | - | 0.042 | - |
|  | $\theta$ [mrad]: | 5.2 | - | 5.0 | - |
|  | $r$ [mm]: | 0.12 | - | 0.06 | - |

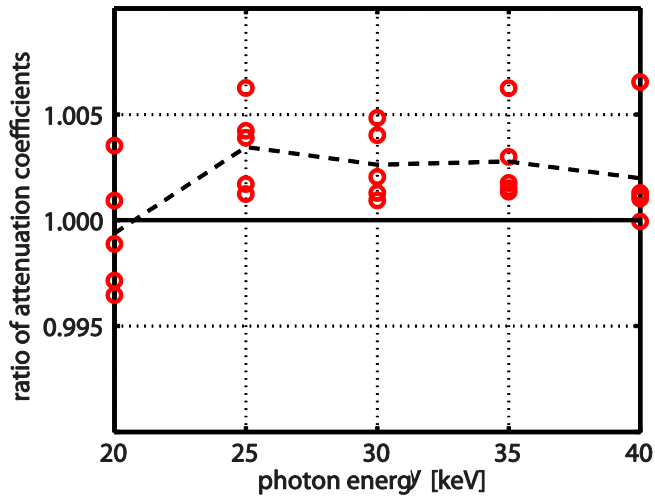

**Figure 3:** Ratio of linear attenuation coefficients determined by spectral measurement and elemental composition for cyst fluid. The markers represent the 5 investigated cyst-fluid samples and the dashed line is the mean.

## 3.2. Spectral measurement results

Figure 4 shows spectral measurement results from all 50 samples of cyst fluid and water in terms of the equivalent thicknesses of PMMA and Al, normalized to a 10-mm sample and a PMMA density of 1.19 g/cm$^3$. Individual measurement points are shaded in grey. The mean and total variability (one standard deviation) are shown as a colored overlay in Figure 4, and are also given in Table 3. The sampled range of the step wedge is shown as a grid around the measurement points, and the angle ($\theta$) and magnitude ($r$) are illustrated for the water Al-PMMA vector. There was a significant difference (at the 1% significance level, determined by two-sample $t$-tests) between cyst fluid and water in terms of Al and PMMA thicknesses, as well as in terms of $\theta$ and $r$. In retrospect it can be noted that the spectral measurements of the 5 samples sent for chemical analysis (Table 2) were not significantly different (at the 5% significance level) from the measurements on all 50 samples (Table 3), which indicates that the subset used for validation was representative.

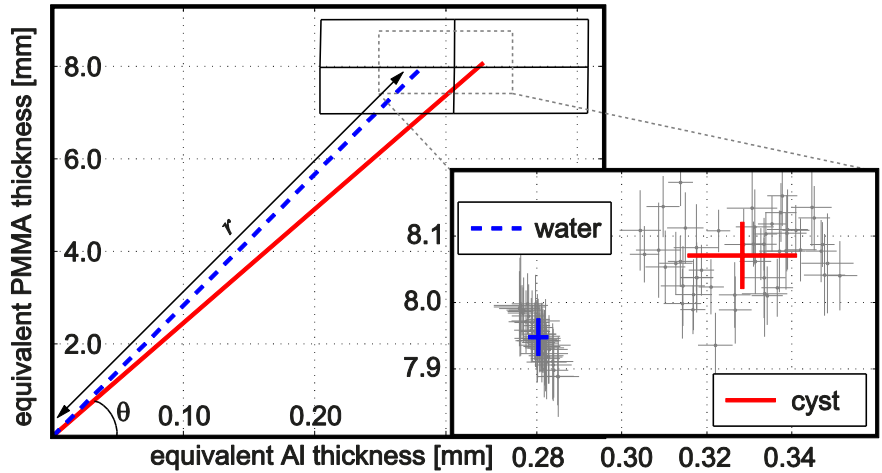

**Figure 4:** Equivalent PMMA and Al thickness measured with the spectral method on 50 10-mm samples of cyst fluid (red solid line) and water (blue dashed line). **Left:** Overview of the Al-PMMA vectors with an illustration of the angle $\theta$ and magnitude $r$. The sampled range of the step wedge is shown as a rectangular grid. **Right:** Close-up of the measurement points. The error bars for individual measurements (shaded) show one standard deviation of the inter-image variability, and the error bars for each sample type (colored) show the total variability.

**Table 3:** Equivalent PMMA and Al thicknesses measured with the spectral method on 50 10-mm samples of cyst fluid and water. All variability measures are quantified as one standard deviation, and the following are given: total variability (total) is the sample-to-sample variability, with the expected quantum noise (total quantum) calculated from all ROIs and all images of each sample. Inter-image variability (inter-image) is the image-to-image variability of the same sample, and intra-image variability is the variability between different ROI positions within the same image. The expected quantum noise (image quantum) is the same for both of these variability measures. Additionally, the angle ($\theta$) and magnitude ($r$) of the Al-PMMA vectors are given and marked in grey to indicate that these measures are not independent of the PMMA and Al thicknesses.

| | | mean | variability | | | | |
|---|---|---|---|---|---|---|---|
| | | | total | total quantum | inter-image | intra-image | image quantum |
| Cyst: | PMMA [mm]: | 8.12 | 0.61% | 0.34% | 1.2% | 1.2% | 1.4% |
| | Al [mm]: | 0.323 | 3.9% | 0.75% | 2.5% | 2.5% | 3.0% |
| | $\theta$ [mrad]: | 39.8 | 4.0% | 0.82% | 2.7% | 2.8% | 3.3% |
| | $r$ [mm]: | 8.12 | 0.61% | 0.34% | 1.2% | 1.2% | 1.4% |
| Water: | PMMA [mm]: | 8.00 | 0.34% | 0.31% | 1.2% | 1.1% | 1.2% |
| | Al [mm]: | 0.276 | 0.82% | 0.80% | 2.9% | 2.7% | 3.1% |
| | $\theta$ [mrad]: | 34.5 | 0.89% | 0.86% | 3.1% | 3.0% | 3.4% |
| | $r$ [mm]: | 8.00 | 0.34% | 0.31% | 1.2% | 1.1% | 1.2% |

For the water measurements, the total variability in the equivalent thicknesses for PMMA and Al was not significantly larger (at the 5% significance level, determined by a chi-square variance test) than the expected quantum noise. The random fluctuations between measurements (inter-image) and the sample variability (intra-image) were also not significantly larger than the expected quantum noise between ROIs. Further, the expected quantum noise between ROIs was approximately 4 times larger than the expected quantum noise between samples. All of these observations are to be expected for homogenous samples and a quantum limited system.

For cyst fluid, the inter- and intra-image variability measures were also not significantly larger than the quantum noise between ROIs, which indicates good sample homogeneity and system stability. The total variability was, however, significantly larger than the expected quantum noise, which is an indication of natural spread among samples.

The linear attenuation coefficients for cyst fluid and water calculated from the spectral measurements at a range of x-ray energies relevant to mammography are shown in Table 4 and plotted in Figure **5**. A close-up of the central region shows the attenuation in more detail and also plots the range of the cyst-fluid measurements as one standard deviation (for the total variability). The energy dependent attenuation of cyst fluid is markedly different from that of water.

**Table 4:** Linear attenuation of cyst fluid and water calculated from the spectral measurements.

| photon energy [keV]: | 15 | 20 | 25 | 30 | 35 | 40 |
|---|---|---|---|---|---|---|
| cyst linear attenuation [cm$^{-1}$]: | 1.76 | 0.852 | 0.533 | 0.391 | 0.318 | 0.277 |
| water linear attenuation [cm$^{-1}$]: | 1.64 | 0.800 | 0.504 | 0.372 | 0.305 | 0.266 |

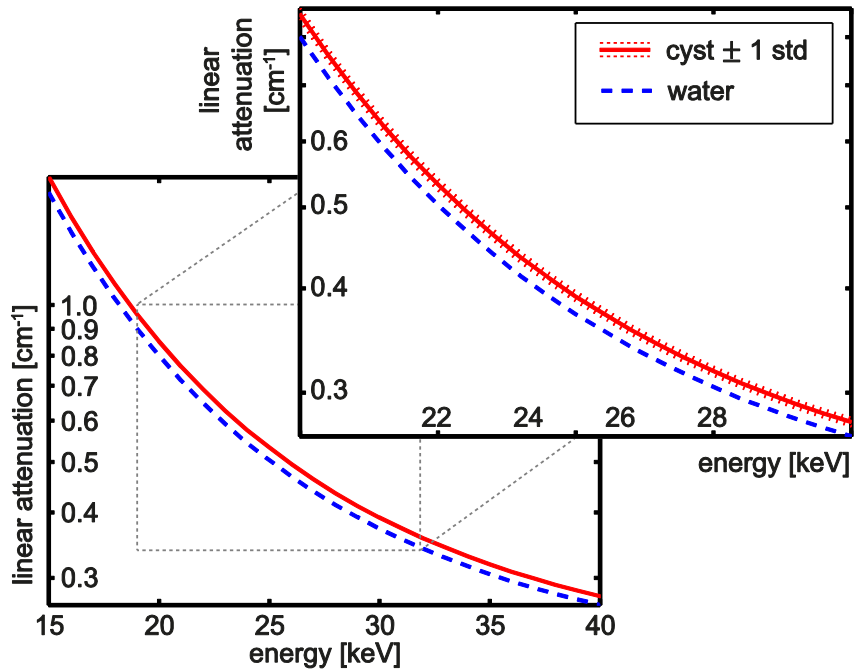

**Figure 5:** Linear attenuation coefficients of cyst fluid and water calculated from the spectral measurements. **Left:** Overview and **Right:** close-up. The dotted lines on either side of the cyst line show ±1 standard deviation of the measured values.

## 4. Conclusions and outlook

There are two main results of this study: 1) A new method to measure x-ray attenuation with a clinical spectral mammography system has been developed, and 2) the linear attenuation coefficient for breast cyst fluid has been determined, to our knowledge for the first time. The equivalent PMMA and Al thicknesses for cyst fluid were significantly different from those for water. There was a relatively large natural spread between different samples of cyst fluid, but the homogeneity of each individual sample was found to be good. The spectral method to measure x-ray attenuation proved to be stable and was validated by also calculating the attenuation from the elemental composition of cyst fluid and water.

The ultimate goal of our study of tissue attenuation is to develop a technique to characterize round lesions as cyst or solid tumor in screening with spectral imaging, and to evaluate the feasibility of the technique. A preliminary version of the technique has been described in a previous publication (Norell *et al*, 2012), and a clinical study has been initiated at the Cambridge Breast Unit. In short, upon detection of a suspicious lesion, the local breast thickness and density (the amount of fibro-glandular tissue) are assessed in a reference region surrounding the lesion using spectral information and knowledge of the attenuation of normal breast tissue. By assuming these two properties to be constant or varying in a predictable way over the lesion, the influence of overlapping tissue can be removed. Subsequently, with knowledge of the attenuation of cysts and tumors, the lesion thickness and contents (cyst or tumor) may be determined.

A current best estimate of the minimum projected lesion size for a confident characterization under the influence of random variations, such as quantum noise and anatomical structures, is 1 cm$^2$ (Norell *et al*, 2012). Systematic uncertainties, including lesion attenuation, will, however, also influence the characterization. In particular, published data on tumor attenuation cover a relatively wide range, which is, on top of natural variability, likely to be at least partly caused by different experimental conditions. We therefore plan to continue this pre-clinical study by measuring tumor attenuation with the same spectral method as for the cyst fluid, and under conditions that are as similar as possible to the screening environment.

Although our main clinical objective is to develop a method to distinguish cysts from solid tumors, the method to measure the x-ray attenuation of tissue samples presented here and the results for cyst fluid will also have more general application in modeling breast imaging with x-rays.

**Acknowledgements**
Part of this work was carried out under the OPTIMAM project supported by the Cancer Research UK & Engineering and Physical Sciences Research Council Cancer Imaging Programme in Surrey, in association with the MRC and Department of Health (England). We are indebted to Paul Hemming of Exeter Analytical UK Ltd, for his advice and assistance with the measurements of chemical composition.